\documentclass[manuscript]{emulateapj}
\usepackage{apjfonts}
\usepackage{epsf}
\usepackage{color}

\newcommand{\HI}{\ion{H}{1}}

\newcommand{\etal}{{et al.}}
\newcommand{\kms}{km s$^{-1}$}
\newcommand{\e}[1]{$\pm$#1}

\def\mlstar{\ifmmode\Upsilon_{\!\!*}\else$\Upsilon_{\!\!*}$\fi}
\def\mlstarR{\ifmmode\Upsilon_{\!\!*}^R\else$\Upsilon_{\!\!*}^R$\fi}

\def\HIm{\ifmmode \hbox{\scriptsize H\kern0.5pt{\footnotesize\sc i}}\else 
H\kern1pt{\small I}\fi}

\shorttitle{Are Dwarf Galaxies Dominated by Dark Matter?}

\shortauthors{Swaters, Sancisi, van Albada, \& van der Hulst}

\submitted{ApJ, accepted}

\begin{document}

\title{Are Dwarf Galaxies Dominated by Dark Matter?}

\author{R. A. Swaters\altaffilmark{1}} \affil{Department of Astronomy,
  University of Maryland, College Park, MD 20742}
\altaffiltext{1}{Present address: National Optical Astronomy
  Observatory, 950 North Cherry Ave., Tucson, AZ 85719}
\email{rob@swaters.net}

\author{R. Sancisi}
\affil{INAF - Osservatorio Astronomico di Bologna, via Ranzani 1, 40127
  Bologna, Italy\\ and \\
Kapteyn Astronomical Institute, University of Groningen, Landleven 12, 
9747 AD Groningen, the Netherlands}

\author{T. S. van Albada, J. M. van der Hulst}
\affil{Kapteyn Astronomical Institute, University of Groningen, Landleven 12, 
9747 AD Groningen, the Netherlands}

\begin{abstract}

  Mass models for a sample of 18 late-type dwarf and low surface
  brightness galaxies show that in almost all cases the contribution
  of the stellar disks to the rotation curves can be scaled to explain
  most of the observed rotation curves out to two or three disk scale
  lengths. The concept of a maximum disk, therefore, appears to work
  as well for these late-type dwarf galaxies as it does for spiral
  galaxies.  Some of the mass-to-light ratios required in our maximum
  disk fits are high, however, up to about 15 in the $R$-band, with
  the highest values occurring in galaxies with the lowest surface
  brightnesses. Equally well-fitting mass models can be obtained with
  much lower mass-to-light ratios.  Regardless of the actual
  contribution of the stellar disk, the fact that the maximum disk can
  explain the inner parts of the observed rotation curves highlights
  the similarity in shapes of the rotation curve of the stellar disk
  and the observed rotation curve. This similarity implies that the
  distribution of the total mass density is closely coupled to that of
  the luminous mass density in the inner parts of late-type dwarf
  galaxies.

\end{abstract}


\section{Introduction}
\label{theintro}

Late-type dwarf galaxies are commonly thought to have slowly rising
rotation curves and to be dominated by dark matter at all radii (e.g.,
Carignan \& Beaulieu 1989; Persic et al.\ 1996; C\^ot\'e et
al.\ 2000). However, in a recent study of a large sample of late-type
dwarf galaxies for which the rotation curves were derived in a uniform
way, taking the effects of beam smearing into account, Swaters et
al.\ (2009) found that the rotation curves of late-type dwarf galaxies
have shapes similar to those of late-type spiral galaxies. For the
dwarf galaxies in their sample, the rotation curves, when expressed in
units of disk scale lengths, rise steeply in the inner parts and start
to flatten at two disk scale lengths, as is usually seen in spiral
galaxies (e.g., Broeils 1992a; Verheijen \& Sancisi 2001). Such a
difference in rotation curve shapes may have implications for the dark
matter properties for late-type dwarf galaxies. We will investigate
the implications for the Swaters et al.\ (2009) sample here.

For spiral galaxies, mass models based on the extended \HI\ rotation
curves indicate that large amounts of dark matter are required to
explain the outer parts of observed rotation curves (e.g., van Albada
\etal\ 1985; Begeman 1987; Broeils 1992a). In most of the galaxies in
these studies, the inner parts of the observed \HI\ rotation curves
(out to two or three disk scale lengths) could be explained by scaling
up the contribution of the stellar disk to the rotation curve, in
agreement with findings based on optical rotation curves (Kalnajs
1983; Kent 1986).  The same scaling, however, leaves large
discrepancies in the outer parts of galaxies with \HI\ rotation curves
(van Albada \& Sancisi 1986).  This discrepancy is interpreted as
evidence for the existence of large amounts of dark matter in
galaxies. Alternatively, the observed discrepancy could be explained
by a different theory of gravitaty, such as MOND (Modified Newtonian
Dynamics; Milgrom 1983; Sanders 1996).

The dark matter properties of galaxies are usually based on mass
modeling of the rotation curves. If the contributions of the visible
components are fixed, then whatever remains is the dark matter. A
major obstacle is that the precise contribution of the stars to the
rotation curve is not known, because the mass-to-light ratio of the
stars is unknown. Upper limits to the mass-to-light ratios have been
obtained by assuming that the contribution of the stellar disk is
maximal (Kalnajs 1983; Kent 1986, 1987; Van Albada and Sancisi
1986). This `maximum disk' solution minimizes the amount of dark
matter required to explain the observed rotation curves. At the same
time, as shown e.g., by van Albada \& Sancisi (1986), the
uncertainties in the stellar mass-to-light ratios allow for a
range in mass models with different dark matter distributions.

Rotation curve studies of the dwarf galaxy DDO~154 (Carignan \&
Freeman 1988; Carignan \& Beaulieu 1989) indicated, however, that this
galaxy is dominated by dark matter at all radii, including the region
well within the optical disk. Even when the contribution of the
stellar disk is scaled as high as is allowed by the observed rotation
curve (i.e., the maximum disk solution), the stellar disk could not be
scaled to explain the observed rotation curves out to two or three
disk scale lengths.  The observations of DDO~154, along with studies
of scaling relations based on relatively few well-studied dwarf
galaxies (e.g., Casertano \& van Gorkom 1991; Broeils 1992a; Persic et
al.\ 1996), led to the generally accepted picture that dwarf galaxies
have slowly rising rotation curves and are dominated by dark matter at
all radii.

There are, however, also studies that provide a different picture, in
which the stellar disks could be scaled to explain all of the inner
rise of the rotation curves (e.g., Carignan 1985; Carignan
\etal\ 1988; Lake \etal\ 1990; Broeils 1992b; Kim \etal\ 1998),
suggesting that the dark matter properties may be similar to
those of spiral galaxies. 

A major problem is that in studies to date the galaxies have been
observed with very different instrumental setups, and that the
rotation curves were derived using different procedures, some of which
may have been prone to introducing systematic errors (see e.g.,
Swaters \etal\ 2002; de Blok \etal\ 2008).  Furthermore, the effects
of beam smearing were not taken into account, even though these can be
important (see e.g., Begeman 1987; Swaters \etal\ 2009).

In order to improve this situation we have obtained \HI\ observations
for a sample of 73 dwarf galaxies with a single instrument (Swaters
1999, hereafter S99; Swaters \etal\ 2002, hereafter Paper~I), as well
as $R$-band observations (Swaters \& Balcells 2002, hereafter
Paper~II). From the \HI\ observations, we derived rotation curves in a
uniform way, taking into account the effects of beam smearing (S99;
Swaters \etal\ 2009, hereafter Paper~III).  From this sample we have
selected 18 high quality \HI\ rotation curves for a detailed mass
model analysis which we report in this paper.

The layout of this paper is as follows. In the next section we will
describe the sample and the rotation curves. In
Section~\ref{themodels} the different components that are used in the
mass models and the fitting of these mass models to the rotation
curves are described. Section~\ref{themodres} presents the results of
the mass modeling. In Section~\ref{thedisc} the results are discussed,
and we present our conclusions in Section~\ref{theconclusions}.

\section{The sample and the rotation curves}
\label{thesample}

The late-type dwarf galaxies in this sample have been observed as part
of the WHISP project (Westerbork HI Survey of Spiral and Irregular
Galaxies; for a more detailed description of the WHISP project and its
goals, see Paper~I).  The galaxies in the WHISP sample have been
selected from the UGC catalog (Nilson 1973), taking all galaxies
with declinations north of $20^\circ$, blue major axis diameters
larger than $1.5'$ and measured flux densities larger than 100
mJy. From this list we selected the late-type dwarf galaxies, defined
as galaxies with Hubble types later than Sd, supplemented with spiral
galaxies of earlier Hubble types but with absolute $B$-band magnitudes
fainter than $-17$. For a detailed description of the selection
criteria, see Paper~I.  Optical $R$-band data for all these galaxies
are presented in Paper~II, which also includes a discussion of the
distance uncertainties for the galaxies presented here.

\begin{deluxetable*}{rrrrrrrrrrrrrrrrr}
\tabletypesize{\scriptsize}
\tablecaption{Optical and dark matter properties\label{tabfits}}
\tablewidth{0pt}
\tablehead{
\colhead{} & \colhead{} & \colhead{} & \colhead{} & \colhead{} & \colhead{} & \colhead{} & \multicolumn{4}{c}{maximum disk} & \multicolumn{3}{c}{minimum disk} & \multicolumn{3}{c}{$\mlstarR=1$} \\
\colhead{} & \colhead{} & \colhead{} & \colhead{} & \colhead{} & \colhead{} & \colhead{} & \multicolumn{4}{c}{\hrulefill} & \multicolumn{3}{c}{\hrulefill} & \multicolumn{3}{c}{\hrulefill} \\
\colhead{UGC} & \colhead{$\mathrm{D}_a$} & \colhead{$M_R$} & \colhead{$h$} & \colhead{$\mu_0^R$} & \colhead{$S_{(2,3)}^h$} & \colhead{$i$} & \colhead{$\mlstarR$} & \colhead{$\rho_0$} & \colhead{$r_c$} & \colhead{$\chi^2_r$} & \colhead{$\rho_0$} & \colhead{$r_c$} & \colhead{$\chi^2_r$} & \colhead{$\rho_0$} & \colhead{$r_c$} & \colhead{$\chi^2_r$} \\
\colhead{(1)} & \colhead{(2)} & \colhead{(3)} & \colhead{(4)} & \colhead{(5)} & \colhead{(6)} & \colhead{(7)} & \colhead{(8)} & \colhead{(9)} & \colhead{(10)} & \colhead{(11)} & \colhead{(12)} & \colhead{(13)} & \colhead{(14)} & \colhead{(15)} & \colhead{(16)} & \colhead{(17)} \\
}
\startdata
  731           &  8.0 &-16.6 &1.65 &23.0 &0.34 &57 &15.1 & 1.6\e0.3 &$\infty$ &0.45& 170\e30& 0.8\e0.1 &0.23 &150\e30 &0.81\e0.09& 0.22\\
 3371\rlap{$^*$}& 12.8 &-17.7 &3.09 &23.3 &0.18 &49 &12.5 & 1.1\e0.2 &$\infty$ &0.73&  37\e6 & 2.1\e0.2 &0.85 & 32\e6  & 2.2\e0.3 & 0.82\\
 4325           & 10.1 &-18.1 &1.63 &21.6 &0.08 &41 & 9.1 &  --      &  --     &1.81& 330\e80& 0.8\e0.1 &1.77 &300\e70 &0.75\e0.10& 1.49\\
 4499           & 13.0 &-17.8 &1.49 &21.5 &0.30 &50 & 2.3 &  14\e5   & 2.8\e0.7&0.58&  65\e15& 1.3\e0.2 &0.15 & 37\e10 & 1.7\e0.3 & 0.22\\
 5414           & 10.0 &-17.6 &1.49 &21.8 &--   &55 & 4.6 &  --      &  --     &1.13&  44\e12& 1.5\e0.3 &0.14 & 30\e8  & 1.7\e0.4 & 0.18\\
 6446           & 12.0 &-18.4 &1.87 &21.4 &0.17 &52 & 4.0 & 5.3\e1.5 & 6.2\e1.6&0.39& 200\e40& 0.8\e0.1 &0.16 &120\e20 &0.95\e0.10& 0.05\\
 7323           &  8.1 &-18.9 &2.20 &21.2 &--   &47 & 3.0 & 2.6\e0.8 &$\infty$ &0.39&  50\e12& 2.0\e0.3 &0.66 & 27\e8  & 2.7\e0.7 & 0.30\\
 7399           &  8.4 &-17.1 &0.79 &20.7 &0.27 &55 & 7.7 &  31\e4   & 3.0\e0.3&1.27& 510\e80& 0.6\e0.1 &0.98 &370\e60 &0.76\e0.08& 0.97\\
 7524\rlap{$^*$}&  3.5 &-18.1 &2.58 &22.2 &0.24 &46 & 7.2 & 1.4\e0.3 &$\infty$ &0.27&  73\e8 & 1.4\e0.1 &0.44 & 59\e7  & 1.5\e0.1 & 0.29\\
 7559           &  3.2 &-13.7 &0.67 &23.8 &0.15 &61 &13.1 & 6.1\e2.0 &$\infty$ &0.22& 100\e40& 0.5\e0.2 &0.09 & 90\e40 &0.55\e0.21& 0.09\\
 7577           &  3.5 &-15.6 &0.84 &22.5 &--   &63 & 0.9 &  --      &  --     &0.40&   8\e4 & 0.8\e0.3 &0.15 &  0     &  --      & 0.63\\
 7603           &  6.8 &-16.9 &0.90 &20.8 &0.54 &78 & 4.1 & 7.9\e2.5 & 4.9\e1.8&0.84& 100\e20& 0.9\e0.1 &0.30 & 64\e18 & 1.2\e0.2 & 0.17\\
 8490           &  4.9 &-17.3 &0.66 &20.5 &0.29 &50 & 4.4 &  35\e10  & 1.9\e0.3&0.39&1100\e200&0.33\e0.04&0.34&570\e150&0.44\e0.05& 0.20\\
 9211           & 12.6 &-16.2 &1.32 &22.6 &0.41 &44 &11.2 & 5.2\e2.2 & 5.1\e2.1&0.58&  86\e26& 1.0\e0.2 &0.31 & 72\e24 & 1.1\e0.2 & 0.24\\
11707\rlap{$^*$}& 15.9 &-18.6 &4.30 &23.1 &0.14 &68 & 9.3 & 0.7\e0.1 &$\infty$ &0.83&  61\e21& 1.7\e0.3 &0.42 & 49\e16 & 1.8\e0.4 & 0.42\\
12060           & 15.7 &-17.9 &1.76 &21.6 &0.07 &40 & 8.3 & 0.5\e0.2 &$\infty$ &0.18&500\e200&0.45\e0.14&0.55 &400\e160&0.48\e0.15& 0.36\\
12632\rlap{$^*$}&  6.9 &-17.1 &2.57 &23.5 &0.16 &46 &15.1 & 1.0\e0.2 &$\infty$ &1.06& 110\e20& 1.0\e0.2 &0.12 &100\e20 & 1.0\e0.2 & 0.13\\
12732\rlap{$^*$}& 13.2 &-18.0 &2.21 &22.4 &0.27 &39 & 7.5 & 2.0\e0.4 & 16\e7   &0.37&  74\e18& 1.4\e0.2 &0.17 & 53\e16 & 1.6\e0.3 & 0.24\\
\enddata
\tablecomments{(1) UGC number; galaxies labeled with a $^*$ have properties similar to LSB galaxies (2) adopted distance in Mpc, from Paper~II, (3) absolute $R$-band magnitude from Paper~II, (4) $R$-band scale length in kpc from Paper~II, (5) extrapolated central $R$-band disk surface brightness, from Paper~II, (6) logarithmic rotation curve shape between 2 and 3 disk scale lengths, from Paper~III, (8) inclination, from Paper~III, (9) maximum disk mass-to-light ratio, in units of M/L$_\odot$, (10,12,14) central halo density, in units of $10^{-3}$ M$_\odot$ pc$^{-3}$, (11,13,15) core radius in kpc.}
\end{deluxetable*}

The sample of late-type dwarf galaxies used here is selected from the
sample of 62 late-type dwarf galaxies for which rotation curves could
be derived in Paper~III. Here, we only selected galaxies with high
quality rotation curves (classified as such in Paper~I) and with
inclinations in the range $39^\circ\le i<80^\circ$ (but mass models
for galaxies with lower-quality rotation curves galaxies were
presented by S99).  The lower limit of $39^\circ$ was chosen to also
include UGC~12732. Note that edge-on galaxies have been excluded,
because their rotation curves are difficult to determine, and also
because the radial distributions of the gas and especially that of the
light from the stellar disk are difficult to measure. We excluded
UGC~11861 because of its high luminosity. The resulting sample, listed
in Table~\ref{tabfits}, contains 18 galaxies.

It is possible that some of the galaxies in our sample are affected by
noncircular motions. However, because the galaxies were selected from
the larger sample presented in Paper~III to be symmetric in nature, it
is unlikely noncircular motions play an important role. Moreover,
because the rotation curves have been derived by averaging over
annuli, effects from noncircular motions are minimized.

As can be seen in Table~\ref{tabfits}, most of the galaxies in our
sample have absolute magnitudes fainter than $M_R=-18$, as expected
for dwarf galaxies. Some are a little brighter than $M_R=-18$, because
the initial sample selection, described in detail in Paper~I, was in
part based on morphological type, and because of uncertainties in the
determination of the absolute magnitudes that were used in the initial
sample selection.

Most of the late-type dwarf galaxies have small exponential disk scale
lengths. The median value for the scale length is 1.6 kpc, and 5
galaxies have scale lengths of less than 1~kpc. However, 7 galaxies
have scale lengths larger than 2 kpc, and 5 of these have surface
brightnesses lower than $\mu_R=22$ mag.  These galaxies have
properties that are similar to those of the class of low surface
brightness (LSB) galaxies studied by e.g., de Blok \etal\ (1995).

The rotation curves have been derived with an interactive procedure,
as described in detail in Paper~III.  First an estimate of each
rotation curve was made based on a simultaneous fit by eye to six
position-velocity diagrams with different position angles.  Next, each
rotation curve was refined by constructing a model data cube, based on
the input rotation curve, and then adjusting the rotation curve to
find the best match between the model cube and the observed data cube.
As the comparison with high-resolution H$\alpha$ data in Paper~III
showed, this procedure made it possible to correct for the effects of
beam smearing to a large degree.

\section{Mass models}
\label{themodels}

The circular velocity is a direct reflection of the gravitational
potential of a galaxy, assuming it is axially symmetric and in
equilibrium:
\begin{equation}
F_r = \frac{\partial\Phi}{\partial r} = -\frac{{v_\mathrm{c}}^2}{r},
\label{eqpot}
\end{equation}
where $F_r$ is the radial force, $\Phi$ the gravitational potential,
$r$ the radius and $v_\mathrm{c}$ the circular velocity. As was
discussed in Paper~III, the corrections for asymmetric drift tend to
be uncertain because the gas pressure usually cannot be determined
accurately. Because the corrections are expected to be small (less
than 1 \kms\ in the inner parts of the rotation curves and less than 3
\kms\ at all radii for 95\% of the galaxies presented here), the
observed rotation curves were used to represent the circular
velocities.

\begin{figure*}[th]
\vspace{-0.1cm}
\resizebox{0.99\hsize}{!}{\includegraphics{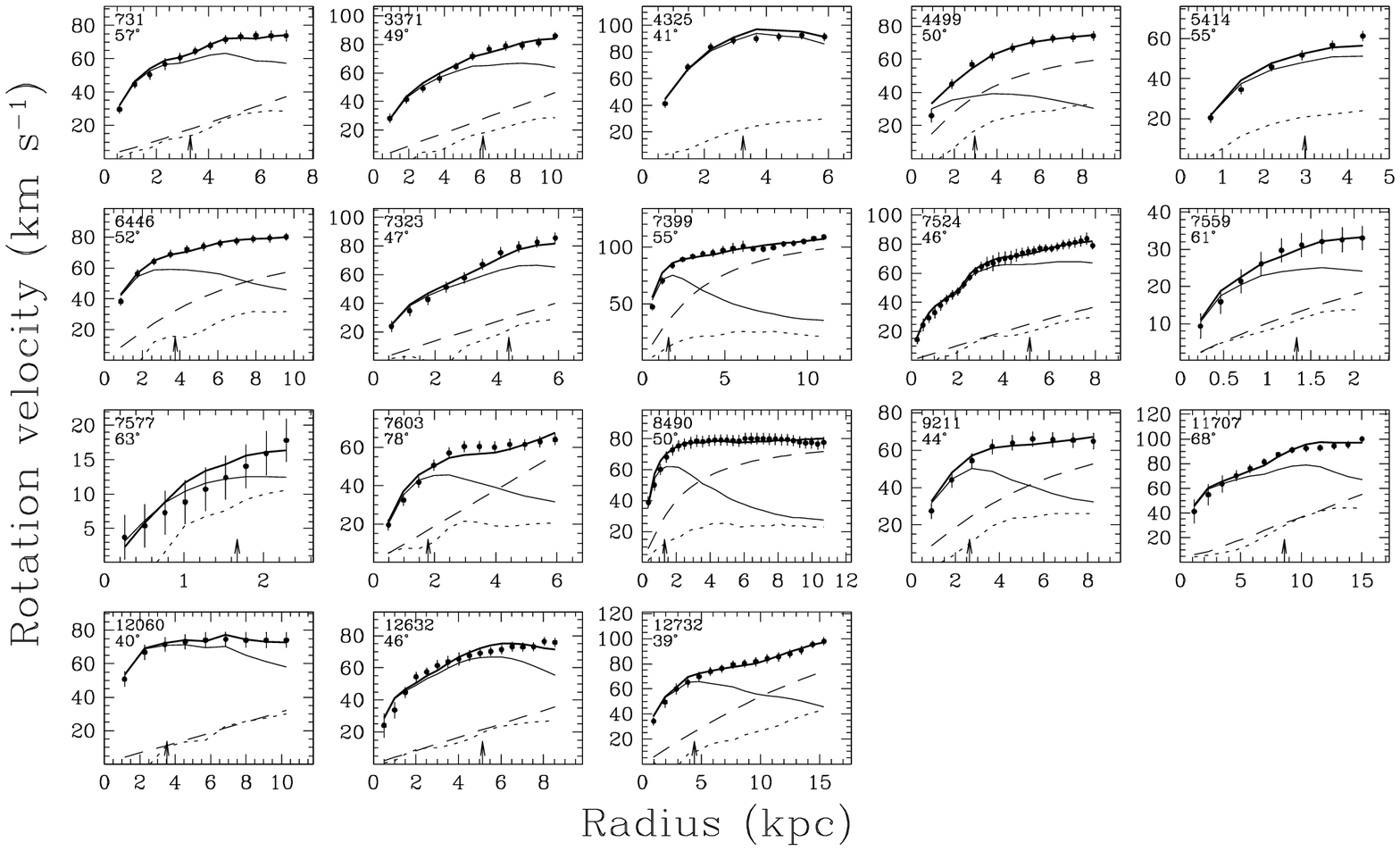}}
\caption{ Maximum disk mass models. The filled circles represent the
  derived rotation curves.  The thin full lines represent the
  contribution of the stellar disks to the rotation curves, the dotted
  lines that of the gas and the dashed lines that of the dark
  halos. The thick solid lines represent the maximum disk mass
  models. The arrows at the bottom of each panel indicate a radius of
  two disk scale lengths.  In the top left corner of each panel the
  UGC number and the inclination are given.}
\label{figallisomax}
\end{figure*}

The gravitational potential is the sum of the
gravitational potentials of the individual mass components in each
galaxy. Expressed in velocities, this sum becomes
\begin{equation}
{v_\mathrm{rot}}^2 = {v_\mathrm{*}}^2 +
{v_\mathrm{g}}^2 + {v_\mathrm{h}}^2,
\label{eqprefit}
\end{equation}
with $v_\mathrm{*}$ the contribution of the stellar disk to the
rotation curve, $v_\mathrm{g}$ the contribution of the gas and
$v_\mathrm{h}$ that of the dark halo. In Equation~\ref{eqprefit} the
contribution of a bulge component is left out because the galaxies in
this sample have little or no bulge. Since we have no prior knowledge
of the stellar mass-to-light ratio, some value \mlstar\ has to be assumed.
The contribution of the gas to the rotation curve includes the
contribution of helium. This scaling of the \HI\ is represented by
$\eta$. Making this explicit, Equation~\ref{eqprefit} becomes
\begin{equation}
{v_\mathrm{rot}} = \sqrt{\mlstar{v_\mathrm{d}}^2 +
\eta{v_\mathrm{\HIm}}^2 + {v_\mathrm{h}}^2},
\label{eqfit}
\end{equation}
where $v_\mathrm{d}$ is the contribution of the stellar disk for a
stellar mass-to-light ratio of unity, and $v_\mathrm{\HIm}$ is that
of the \HI\ only. Each of the individual components in this equation
is described below.

\subsection{The contribution of the stellar disk}

The $R$-band luminosity profiles presented in Paper~II have been used
to calculate the contribution of the stellar disk to the observed
rotation curve, assuming the mass-to-light ratio \mlstarR\ is
independent of radius.  Because most of the galaxies in this sample
have light profiles that are well represented by an exponential disk,
and generally have little or no bulge, the light profiles were not
decomposed into a disk and a bulge component.  The contribution of the
stellar disk to the rotation curve was calculated using the
prescription given in Casertano (1983), assuming that the galaxies are
optically thin and assuming an intrinsic thickness of $q_0=0.2$. This
value was chosen because it appears to be a suitable value for the
average of the intrinsic thickness over the range of galaxy types
included in this sample.  Late-type disk galaxies, of morphological
types around Sd, are reported to have intrinsic thicknesses closer to
$q_0=0.1$ (e.g., de Grijs 1997; Schwarzkopf \& Dettmar 2000; Bizyaev
\& Kajsin 2004), but for late-type dwarf galaxies higher values may be
found, up to $q_0=0.4$, especially towards the lowest luminosity
systems (van den Bergh 1988; Binggeli \& Popescu 1995; Sung
\etal\ 1998). We estimate the uncertainty in the amplitude of the
rotation curve introduced by the variations in $q_0$ from galaxy to
galaxy to be $\sim5\%$ (see S99 for more details).

The major uncertainty in determining a mass model for a galaxy is the
mass to light ratio \mlstar, which is not known a priori and cannot be
derived from the rotation curve fits alone. Equally well fitting mass
models can be obtained with a range in \mlstar\ (e.g. van Albada
\etal\ 1985; Swaters \etal\ 2000, 2003; Dutton \etal\ 2005; see also
Section~\ref{themodres}). However, it is possible to obtain limits on
\mlstar\ by scaling up the contribution of the stellar disk to explain
most of the observed rotation curve in the inner parts (the so-called
maximum disk hypothesis), and by reducing the contribution of the
stellar disk to its minimum while still obtaining a good fit to the
rotation curve.

\begin{figure*}[th]
\vspace{-0.1cm}
\resizebox{0.99\hsize}{!}{\includegraphics{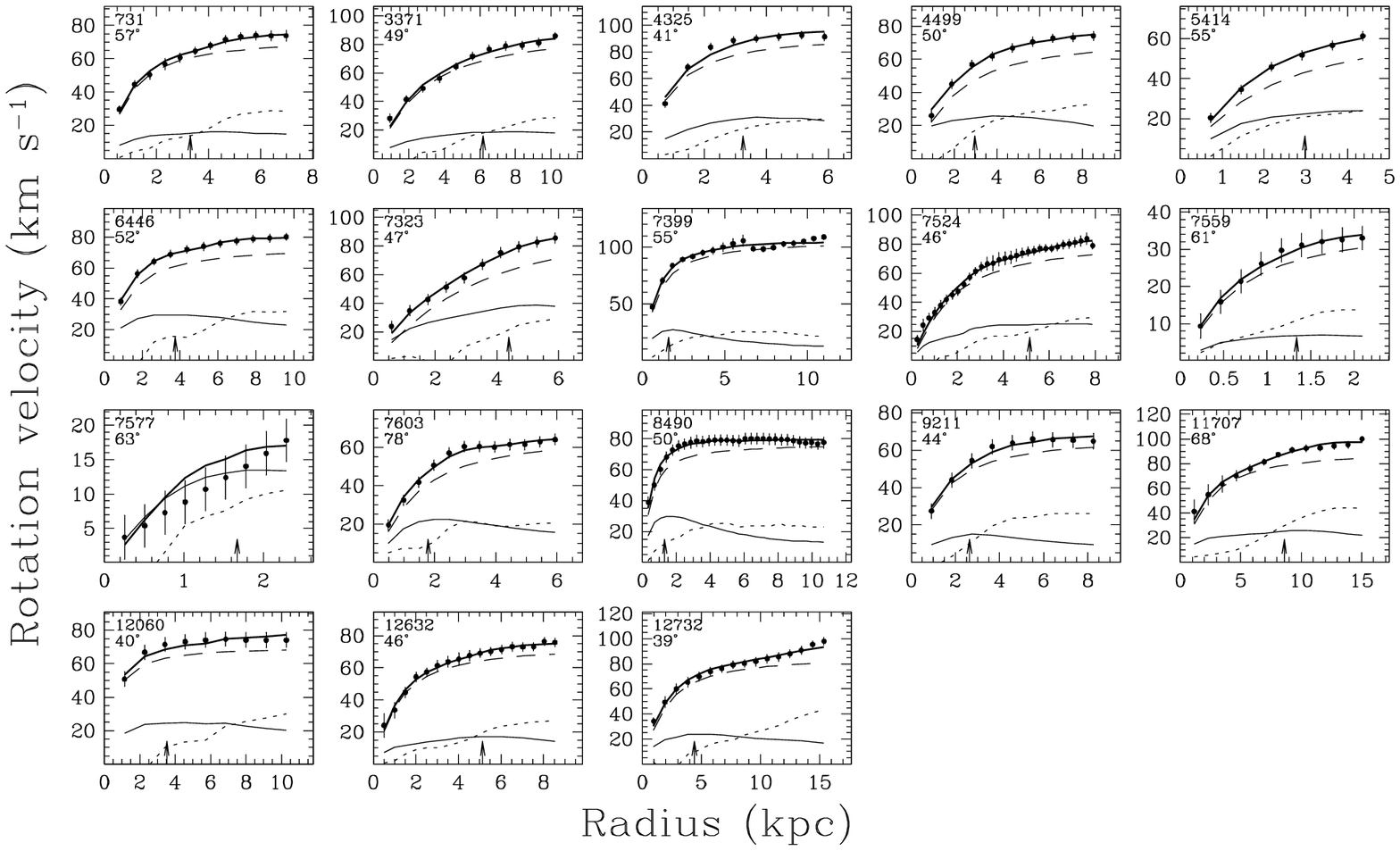}}
\caption{ Same as Figure~\ref{figallisomax}, but for the $\mlstarR=1$ models.}
\label{figallisoml1}
\end{figure*}

\subsection{The contribution of the gas}

The procedure for deriving the contribution of the gas to the rotation
curve is very similar to that of the stellar disk. The \HI\ radial
profiles presented in Paper~I were used to calculate this
contribution. We assumed that the gas layer has an intrinsic thickness
of $q_0=0.2$, the same value that was assumed for the thickness of the
stellar disk. Little is known about the vertical distribution of the
gas in late-type dwarf galaxies. There is some evidence that the gas
and the stars have the same vertical distribution (Bottema
\etal\ 1986). Fortunately the precise choice of the thickness of the
\HI\ layer has very little influence on the shape or amplitude of the
rotation curve and the uncertainty is smaller that that introduced by
assuming a constant thickness of $q_0=0.2$ for the optical disk.

In order to derive the contribution of the gas to the rotation curve,
the \HI\ was assumed to be optically thin. To correct for the mass
fraction of helium, the \HI\ mass was scaled by a factor $\eta=1.32$,
derived assuming a primordial Helium fraction of 24\% (Steigman 2007).

Other gas components that might contribute to the rotation curve, such
as molecular hydrogen, have been ignored. To date, most evidence
suggests there is much less molecular hydrogen in late-type dwarf
galaxies than there is \HI, even when taking into account that the
conversion factor to convert the observed CO flux into a molecular gas
column density may be substantially higher in late-type dwarf galaxies
(e.g., Israel \etal\ 1995; Boselli \etal\ 2002; Buyle \etal\ 2006).
Moreover, if molecular gas were to have the same radial distribution
as the stellar disk, its contribution is implicitly included in the
maximum disk fits.

\subsection{The contribution of the dark halo}

Several different halo radial mass profiles have been proposed over the
years, most of which produce good fits to the observed rotation
curves. Halo mass-density profiles as predicted by cosmological
simulations have received a great deal of attention, because a
confrontation of these profiles against the observed rotation curves
provides a powerful test of these cosmological models. For most of the
sample presented here, such tests have been presented in van den Bosch
\& Swaters (2001) and Swaters et al.\ (2003).

In this paper, we focus on an empirical description of the properties
of the dark matter halos. For this purpose, we have used the density
profile of an isothermal sphere that has often been used in the
literature to represent the dark halo. The advantage is that it will
be easier to compare our results with those of other studies (e.g.,
Broeils 1992a; Verheijen 1997; Noordermeer 2006; see also
Section~\ref{subcomp}).

The radial density profile of an isothermal sphere can be approximated by
\begin{equation}
\rho(r)=\rho_0\left[1+\left(\!\frac{r}{\,r_c}\right)^2\:\right]^{-1},
\label{eqisorad}
\end{equation}
where $\rho_0$ is the central density of the halo, and $r_c$ the core
radius. This density profile gives rise to a rotation curve of the
following form
\begin{equation}
v_h(r)=\sqrt{4\pi\mathrm{G}\rho_0r_c^{\,2}\left[1-\frac{r_c}{\!r}
\mathrm{arctan}\left(\!\frac{r}{\,r_c}\right)\right]}.
\label{eqisovrot}
\end{equation}

\begin{figure*}[th]
\vspace{-0.1cm}
\resizebox{0.99\hsize}{!}{\includegraphics{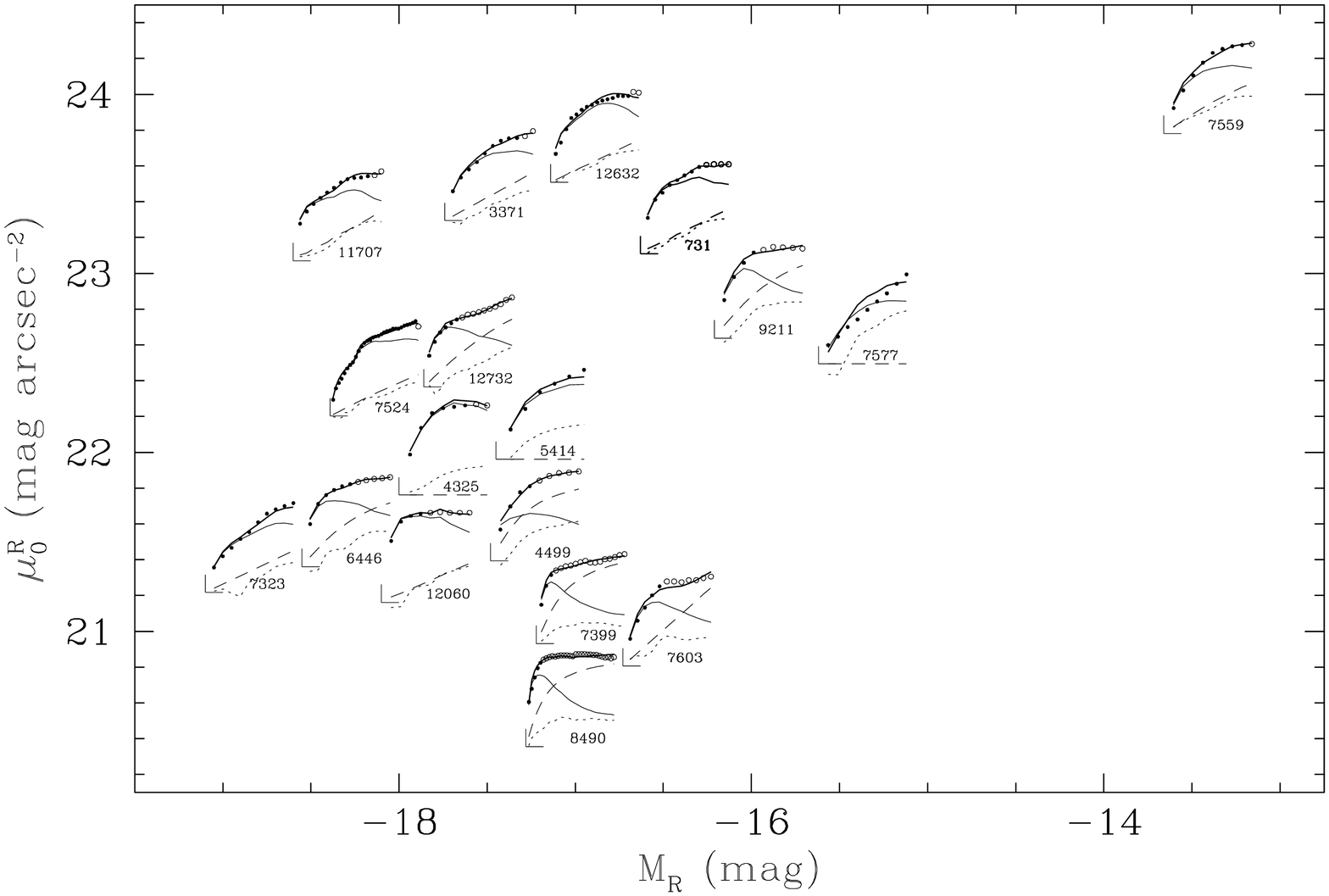}}
\caption{ Rotation curves and the fitted maximum disk mass models
  plotted as a function of absolute magnitude $M_R$ and central disk
  surface brightness $\mu_0^R$. The origin of each rotation curve
  indicates the corresponding $M_R$ and $\mu_R$ of that galaxy. Where
  necessary, the rotation curves have been shifted slightly to avoid
  overlap between the different rotation curves. The rotation curve
  data are represented by circles, filled circles for radii smaller
  than three disk scale lengths, open circles for large radii. Coding
  of the lines is the same as in Figure~\ref{figallisomax}.  The
  rotation curves have been scaled by maximum velocity and radius of
  the last measured point. Note that the relative contribution of the
  stellar disk to the rotation curve does not appear to depend on
  absolute magnitude or surface brightness.}
\label{figisomaxplot}
\end{figure*}

\subsection{Rotation curve fits}
\label{subrotcurfits}

The relative contribution of each of the three components described
above is determined by a simultaneous fit of the right-hand side of
Equation~\ref{eqfit} to the observed rotation curve. For an isothermal
halo, there are four free parameters in Equation~\ref{eqfit}.
However, the contribution of the atomic gas to the rotation curve is
known, and therefore, with the correction for the mass fraction of
helium, $\eta$ is fixed to 1.32. On the other hand, the value of
\mlstar\ is not known beforehand, and a range of \mlstar\ and halo
parameters exists which produce a good fit to the data (van Albada
\etal\ 1985).  Mathematically, it is possible to derive the parameters
for the isothermal halo and the value of \mlstar\ from a best fit to
the observed rotation curve. However, the physical meaning of such a
best fit is unclear, as the values found from such a fit depend on the
small scale variations and uncertainties in the rotation
curves. Because of these uncertainties in \mlstar, two limiting cases
are presented. In one case, the contribution of the stellar disk is
scaled up to explain most of the rotation curve out to two or three
disk scale lengths. This is the maximum disk solution.  In the other
limiting case, the contribution of the stellar disk is completely
ignored, and the rotation curve is fitted with the contributions of
dark halo and gas only.

The maximum disk fits were done by eye, ensuring that the rotation
curve of the stellar disk contributes maximally to the observed
rotation curve within its observational uncertainties, following e.g.,
Broeils (1992a), Verheijen (1997), and Noordermeer (2006). We have
refit several mass models from the literature using our methods. We
were able to reproduce the original results to better than 10
percent. A mass model is only considered to be compatible with a
maximum disk if the stellar disk can be scaled up to explain at least
$\sim 80$\% of the rotation velocity at 2.2 disk scale lengths (e.g.,
Sackett 1997; see also Section~\ref{themain}).

There has been considerable effort over the past decade to relate the
broad-band colors of galaxies to their \mlstar\ through stellar
population synthesis modeling (e.g., Bell \& de Jong 2001; Bell
\etal\ 2003; Zibetti \etal\ 2009). The advantage of using such models
is that it provides an independent estimate of \mlstar\ and possibly a
more realistic value of \mlstar.  However, there is some uncertainty
in \mlstar\ as derived from population synthesis modeling, because of,
for example, uncertainties in the star formation history, the initial
stellar mass function, metallicity, and the details of the late-phases
of stellar evolution. The combined uncertainty may be a factor 4 or
even larger if some of uncertainties are systematic (see e.g.,
Bershady \etal\ 2010).

The six galaxies in our sample with both $B$ and $R$-band
observations presented in Paper~II have an average $B-R$ color of
0.8~mag. Following Bell \etal\ (2003), this corresponds in
$\mlstarR\sim 0.8$, although this value of \mlstarR\ is uncertain, as
discussed above.  However, for any \mlstarR\ near unity, all but one
of the galaxies in our sample are completely dominated by dark matter,
so the exact choice of \mlstarR\ has little influence on the derived
dark halo parameters (see also Section~\ref{themodres}).  Given these
considerations, and because we had already calculated the contribution
of the stellar disk for $\mlstarR=1$ for the input of the mass
modeling, we used these to determine the dark halo properties in these
population synthesis-based restricted-\mlstar\ fits.

The rotation curve fits are shown in Figures~\ref{figallisomax} and
\ref{figallisoml1}. The minimum disk fits are not shown separately
because in most cases the mass models are very similar to those for
$\mlstarR=1$. The fit parameters are given in Table~\ref{tabfits}.
The uncertainties on the fitted parameters have been derived from the
68\% confidence levels and include the covariance between $\rho_0$ and
$r_c$. The uncertainties on the derived rotation velocities are
non-Gaussian, as the points in the rotation curves are correlated, and
the rotation curves and their uncertainties can be affected by
systematic effects. The confidence levels and the corresponding
uncertainties should be considered estimates.

\begin{figure}[ht]
\resizebox{\hsize}{!}{\includegraphics{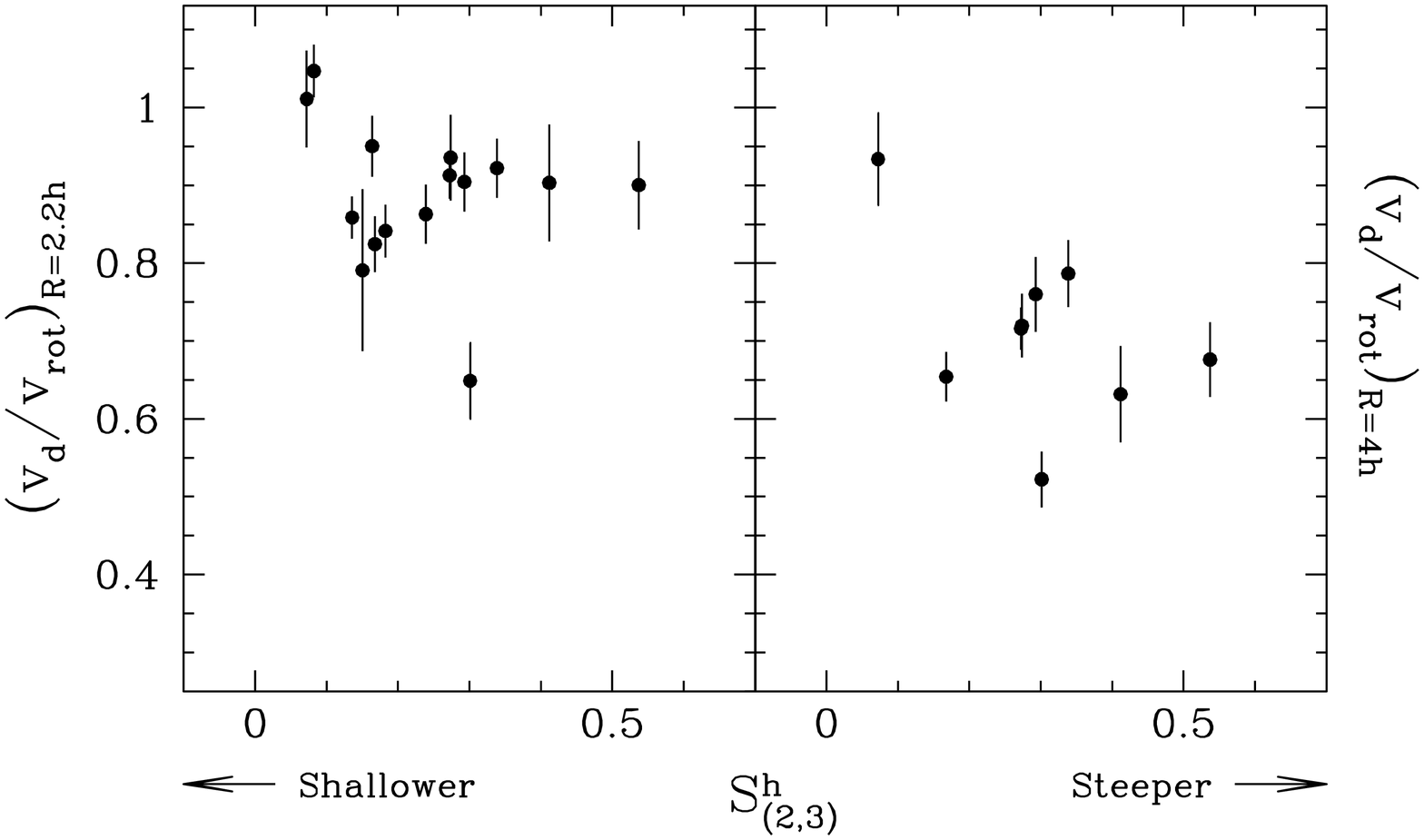}}
\caption{ Relative contribution of the stellar disk to the rotation
  curve for the maximum disk fits at 2.2 disk scale lengths (left
  panel) and at 4 disk scale lengths (right panel) versus logarithmic
  slope between two and three disk scale lengths $S_{(2,3)}^h$.}
\label{figrelvds23h}
\vspace{0.75cm}
\resizebox{\hsize}{!}{\includegraphics{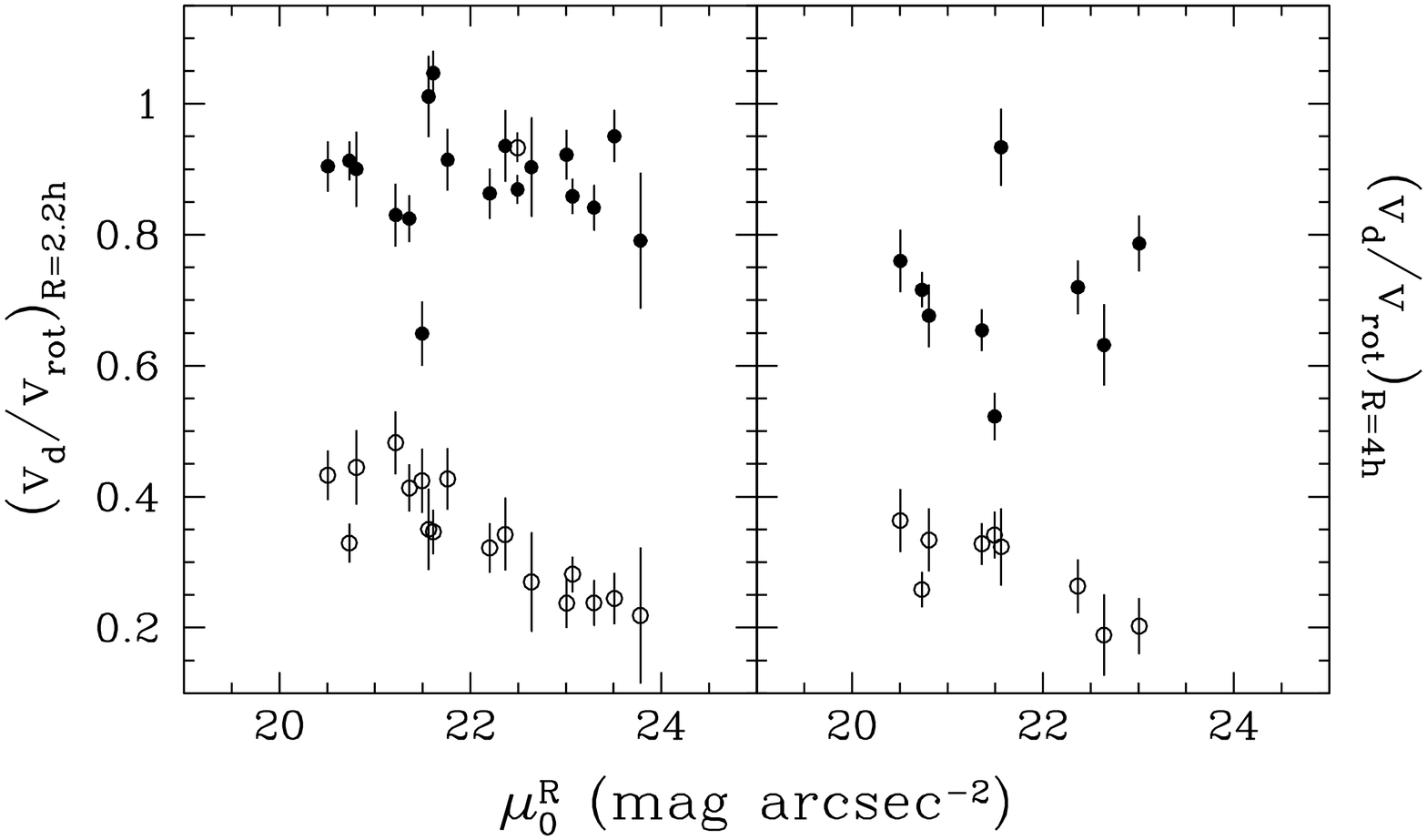}}
\caption{ Relative contribution of the stellar disk to the rotation
  curve for the maximum disk fits (filled circled) and fits with
    $\mlstarR=1$ (open circles) at 2.2 disk scale lengths (left
  panel) and at 4 disk scale lengths (right panel) versus central disk
  surface brightness $\mu_0^R$.}
\label{figrelvdmu}
\end{figure}

\section{Results from fitting mass models}
\label{themodres}

In this section we will present the results from fitting the different
mass models: maximum disk, minimum disk, and fixed \mlstarR.

\subsection{Maximum disk fits}
\label{submaxdisk}

The most striking result from the maximum disk fits, presented in
Figure~\ref{figallisomax}, is that in all 18 galaxies the contribution
of the stellar disk can be scaled up to explain most of the inner
parts of the rotation curves out to two or three disk scale lengths,
with the exception of UGC~4499.  Thus, it seems that the fact that the
stellar disk can be scaled to explain most of the inner parts of the
rotation curves of late-type dwarf galaxies is rule rather than
exception. This is in contrast with the results from several previous
studies of dwarf galaxies, in which the stellar disk only can explain
a small fraction of the inner rise; this is examined in more detail in
Section~\ref{subcomp}.

To illustrate how the maximum disk mass models depend on the optical
galaxy properties, we have plotted the mass models against surface
brightness and absolute magnitude in Figure~\ref{figisomaxplot}. In this
figure, the origin of each model indicates the $\mu_R$ and $M_R$ for
each galaxy. The rotation curves are represented by filled circles for
radii smaller than three disk scale lengths, and by open circles for
larger radii.  From an inspection of Figure~\ref{figisomaxplot} it
appears that there is no clear trend between the relative contribution
of the stellar disk to the rotation curve on the one hand, and
rotation curve shape, luminosity or surface brightness on the
other. Maximum disks are seen at all surface brightnesses and all
absolute magnitudes.

This point is demonstrated in more detail in
Figures~\ref{figrelvds23h} and~\ref{figrelvdmu}.
Figure~\ref{figrelvds23h} shows that there is no clear correlation
between the rotation curve shape as determined by the logarithmic
slope between two and three disk scale lengths $S_{(2,3)}^h$, and the
relative contribution of the stellar disk to the rotation curve at 2.2
disk scale lengths (see Paper~III for details). Independent of the
rotation curve shape, the contribution of the stellar disk to the
rotation curve is about 90\% at 2.2 disk scale lengths.
Figure~\ref{figrelvds23h} also shows the correlation between rotation
curve shape and the relative contribution of the stellar disk at four
disk scale lengths. The contribution of the stellar disk to the
rotation curve has dropped to about 70\%. There is one galaxy,
UGC~12060, where even at four disk scale lengths the stellar disk can
still explain most of the observed rotation curve.  This galaxy has
excess light between three and four disk scale lengths, when compared
to an exponential fall-off.

From Figure~\ref{figrelvdmu} it is clear that the maximum disk
hypothesis works equally well for all surface brightnesses.

From the results presented here, it is apparent that for the vast
majority of these late-type dwarf galaxies the contributions of their
stellar disks to the rotation curves can be scaled to explain most of
the inner parts of the observed rotation curves.  Therefore, these
dwarf galaxies do not necessarily have insignificant mass in their
stellar disks and dominant dark halos, as concluded in earlier work.
For the maximum disk fits, dark matter only becomes important at radii
larger than three or four disk scale lengths. These maximum disk mass
models for late-type dwarf galaxies are similar to those seen in
spiral galaxies (e.g., Begeman 1987; Broeils 1992a; Verheijen 1997).

\subsubsection{Stellar mass-to-light ratios}

\begin{figure}[ht]
\resizebox{\hsize}{!}{\includegraphics{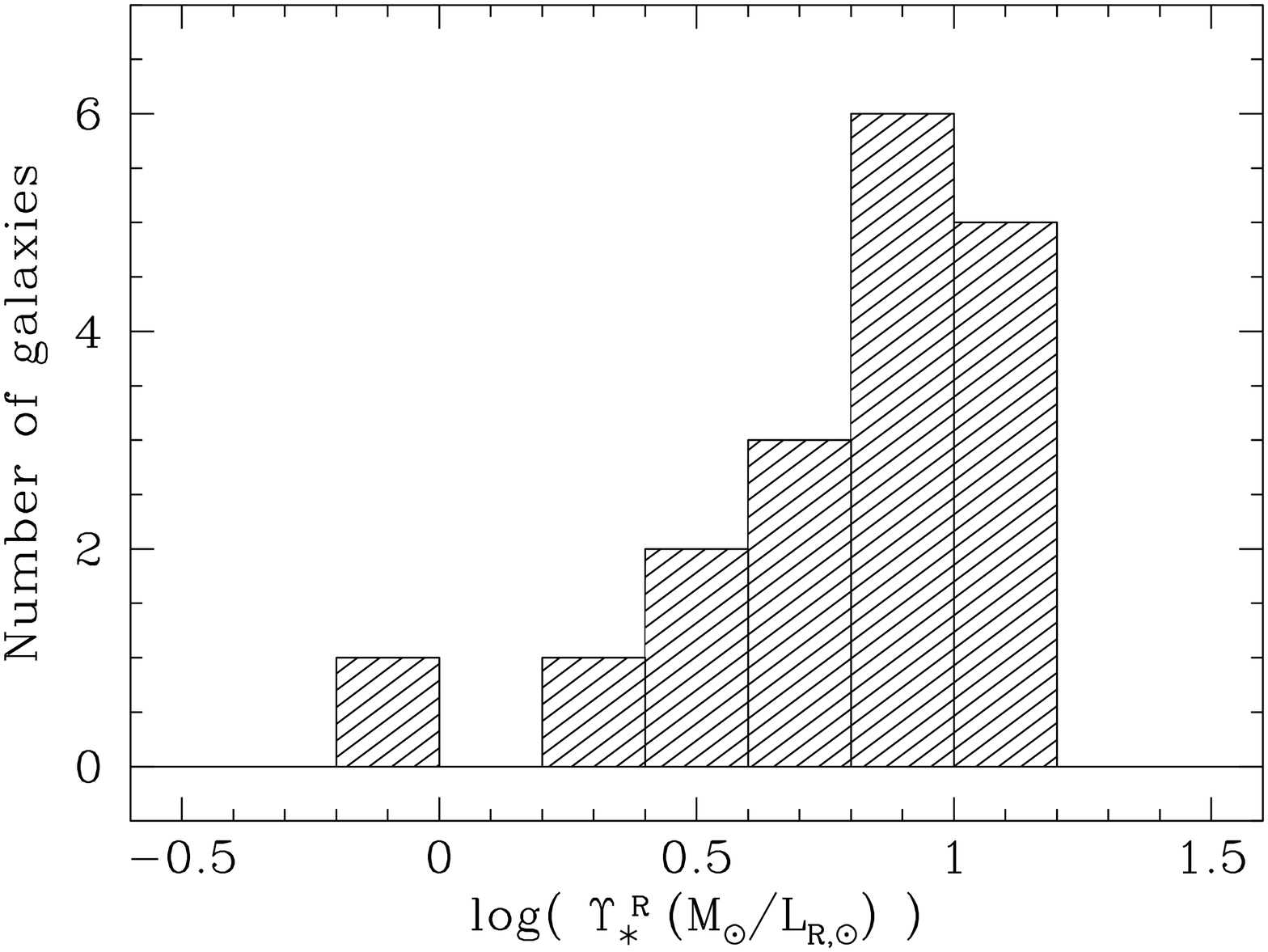}}
\caption{ Distribution of maximum disk $R$-band mass to light ratios
  \mlstarR. }
\label{fighistml}
\vspace{0.75cm}
\resizebox{\hsize}{!}{\includegraphics{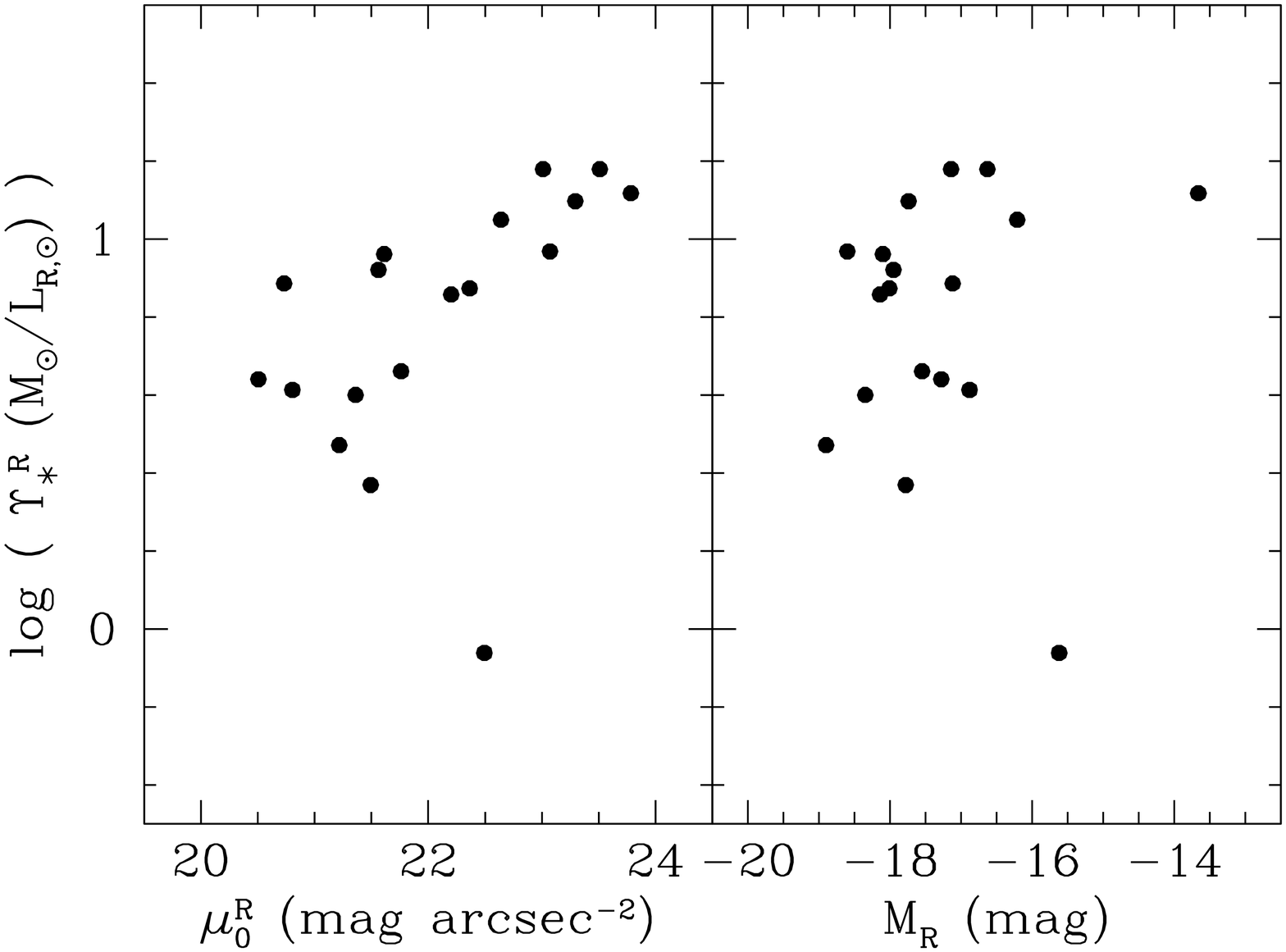}}
\caption{ $R$-band stellar mass-to-light ratios \mlstarR\ as
  determined from the maximum disk fits versus central disk surface
  brightness $\mu_0^R$ (left panel) and versus absolute $R$-band
  magnitude $M_R$ (right panel). }
\label{figmlvsmuabs}
\end{figure}

Although for the majority of the late-type dwarf galaxies in
our sample the contribution of the stellar disk can explain most of
the rotation curve in the inner parts, the stellar mass-to-light
ratios \mlstarR\ may be high.  In Figure~\ref{fighistml} a histogram
is presented of the derived values of \mlstarR\ for the maximum disk
fits. The average value for \mlstarR\ is 7.7, but ranges from 1 to 15,
with the high end being well outside the range predicted by current
population synthesis models. These high values of \mlstarR\ will be
discussed in more detail in Section~\ref{thedisc}.

Figure~\ref{figmlvsmuabs} shows \mlstarR\ plotted against the central
surface brightness $\mu_R$ and the absolute magnitude $M_R$.  Galaxies
with lower surface brightnesses have higher values for \mlstarR\ under
the assumption of maximum disk (see also section~\ref{secmls}).

\subsubsection{Isothermal halo properties}

In three cases (UGC~4325, UGC~5414 and UGC~7577), the rotation curve
can be explained by the stellar disk and the \HI\ alone; see
Figure~\ref{figallisomax}.  Hence, the fitted dark halos will have
central densities close to zero, and the core radii are not
constrained. For several galaxies with more extended rotation curves
(see Figure~\ref{figallisoml1}), the rotation curves of the dark halos
have a solid body appearance, indicating that the dark matter halos
have constant density cores.  In these cases, the core radius is also
unconstrained and only the halo density can be determined. Only for
eight galaxies with sufficiently extended rotation curves both the
core radius and the central density can be determined reliably.

\begin{figure}
\resizebox{\hsize}{!}{\includegraphics{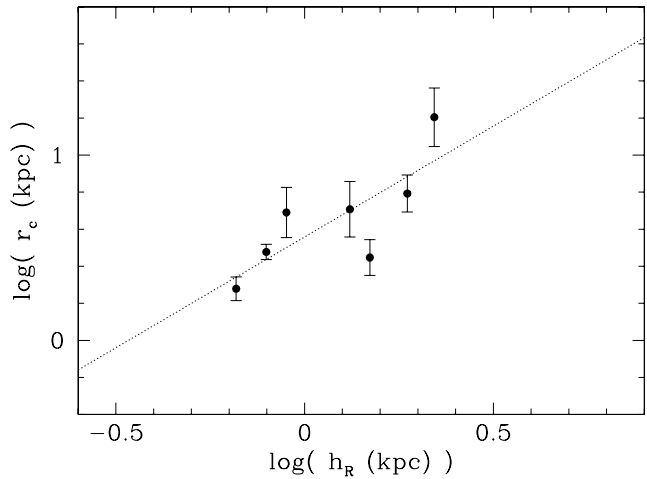}}
\caption{ Halo core radius versus $R$-band optical disk scale length
  for the maximum disk fits, for those galaxies for which the core
  radius could be reliably determined.}
\label{figrcvsh}
\end{figure}

Figure~\ref{figrcvsh} shows that the core radius $r_c$ and the disk
scale length $h$ are correlated.  The halo core radius is on average
$3.5\pm 0.5$ times as large as the disk scale length.  A similar ratio
of $3.0\pm 0.5$ is found for the spiral galaxies presented in Broeils
(1992a), demonstrating the similarity between the maximum disk fits of
the late-type dwarf galaxies in our sample and those of spiral
galaxies. Such a ratio of core radius to disk scale length of about
three is expected for galaxies with a more or less flat rotation
curve, as was also mentioned by van Albada \& Sancisi (1986), who
found that the expected ratio is about 3.5.

For scaling relations between the optical properties of the galaxies
in our sample and the relative fractions of mass in stars, gas, and
dark matter, we refer to S99.

Note from Figure~\ref{figallisomax} that in the case of maximum disk
models the shapes of the halo rotation curves often are similar to the
shapes of the gas rotation curves. This is remarkable and indicates
that the ratio of dark matter surface density to gas surface density
is constant in the outer parts. This was noted for spiral galaxies by
Bosma (1981) and further investigated by Hoekstra et al. (2001). A
detailed study of the apparent link between the \HI\ and dark matter
densities will be the subject of an upcoming paper.

~

\subsection{Fits with fixed mass-to-light ratios}
\label{secmlfixed}

A lower limit to the contribution of the stellar disk to the rotation
curve can be obtained by decreasing \mlstarR\ to its lowest possible
value while the mass model still produces a good fit to the rotation
curve. For the late-type dwarf galaxies in this sample, the rotation
curves can be well fitted using only a dark halo and the \HI\ (i.e.,
$\mlstarR=0$). A minimum disk fit thus does not provide a useful lower
limit on \mlstarR, but it does provide an upper limit on the halo mass
within a given radius and on the central density, and a lower limit on
the halo core radius. The results for the minimum disk fits are given
in Table~\ref{tabfits}. The mass models themselves are not shown
because the dark halos are very similar to those for $\mlstarR=1$,
shown in Figure~\ref{figallisoml1}. Details of the correlations
between the minimum-disk dark halo properties and optical properties
were presented in S99.

As was shown above, and as will be discussed in more detail in
Section~\ref{thedisc}, the maximum-disk \mlstarR\ values are higher
than expected from population synthesis models. To investigate the
dark halo properties for values of \mlstarR\ that are in agreement
with population synthesis models (as was described in
Section~\ref{subrotcurfits}), we also fit mass models for
$\mlstarR=1$.  The results of the fits are presented in
Table~\ref{tabfits}, and the fits themselves are shown in
Figure~\ref{figallisoml1}.

With $\mlstarR=1$, the relative contribution of the stellar disk to
the rotation curve is tightly correlated with surface brightness, as
can be seen in Figure~\ref{figrelvdmu}. Such a tight correlation is
expected for models with a fixed $\mlstarR$, because in that case the
surface brightness is directly linked to the surface mass density and
hence to the relative contribution of the stellar disk.

\section{Discussion}
\label{thedisc}

\subsection{Main result}
\label{themain}

The main result of the present study is that the contribution of the
stellar disk can be scaled up (``maximum disk'') to explain the inner
parts of the rotation curve of all but one of the 18 late-type
galaxies in the present sample.  These maximum disk models are similar
to those for late-type spiral galaxies in which the stellar disks can
also be scaled to explain the inner rotation curves (e.g., Kalnajs
1983; Kent 1987; Begeman 1987; Broeils 1992a; Verheijen 1997) and also
those of the LSB galaxies presented in Swaters \etal\ (2000; 2003).

\begin{figure}
\resizebox{\hsize}{!}{\includegraphics{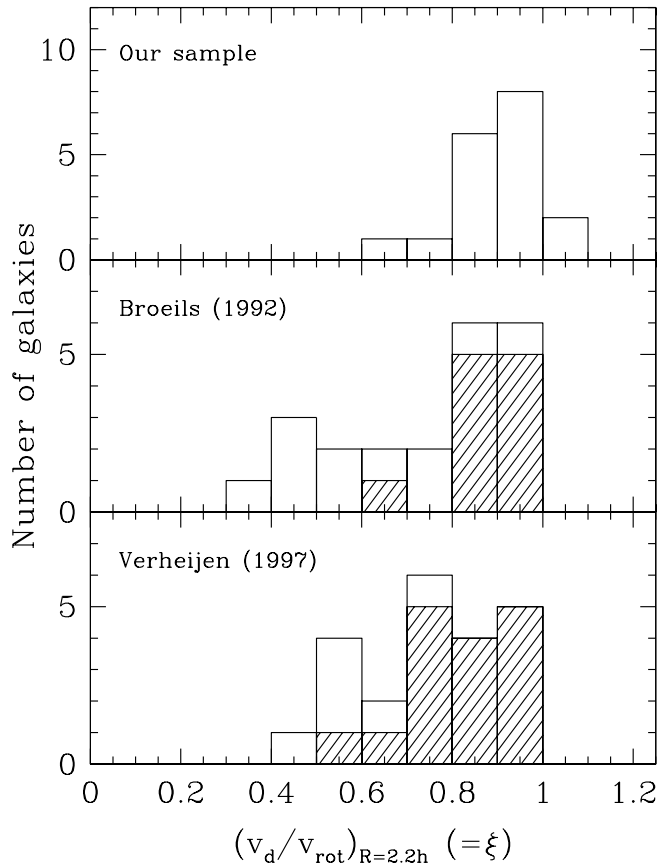}}
\caption{ Distribution of the fractional contribution of the maximum
  stellar disk to the rotation curve at 2.2 disk scale lengths for the
  galaxies in our sample, and the samples presented in Broeils (1992a)
  and Verheijen (1997).  The shaded areas represent galaxies with
  numeric Hubble types earlier than Sd, the open areas represent
  galaxies with Hubble type Sd or later.}
\label{figmaxdiskfrac}
\end{figure}

How well maximum disk works is described by $\xi$ the fractional
contribution of the maximum stellar disk to the rotation curve at two
disk scale lengths. A low value for $\xi$ means that even when scaled
as high as possible, the stellar disk can only explain a small
fraction of the observed rotation curve. A high value of $\xi$ (near
unity) means that the stellar disk can be scaled to explain most or
all of the observed rotation curve.

For the galaxies in our sample $\xi=0.88$, with a dispersion around
the mean of $0.09$. This is nearly identical to the average for the
spiral galaxies in the sample of Broeils (1992a), for which $\xi=0.86
\pm 0.07$, and to the average for spiral galaxies in the Verheijen
(1997) sample, with $\xi=0.81 \pm 0.12$ (see
Figure~\ref{figmaxdiskfrac}). 

If we assume that the values for $\xi$ near unity mean that these
dwarf galaxies physically have maximal disks (despite the high
required mass-to-light ratios), then the stellar disk can explain the
mass distribution over the optical parts of the galaxy, and dark
matter only becomes relevant at large radii.

Under the same assumption, the average ratio of stellar-to-dark
mass within four disk scale lengths is $1.0\pm 0.6$ for the dwarf
galaxies in our sample.  For the spiral galaxies in the sample
presented in Broeils (1992a) the average stellar-to-dark mass ratio
within four disk scale lengths is $0.9\pm 0.4$. Thus, the maximum disk
fits for dwarf galaxies and spiral galaxies are very similar, with
similar relative amounts of mass in the dark and luminous components
within similar numbers of disk scale lengths. In these maximum disk
fits, dwarf galaxies are not dominated by dark matter at all
radii. Dark matter is only needed to explain the outer part of the
rotation curve.

\subsection{Comparison with previous work}
\label{subcomp}

The conclusion reached here contrasts with the general notion from
previous work that in late-type dwarf galaxies dark matter dominated
everywhere.  The average value of $\xi$ in the comparison samples of
Broeils (1992a) and Verheijen (1997) when including only late-type
dwarf galaxies is lower than what we find here.  For the late-type
dwarf galaxies in the sample of Broeils (1992a), $\xi=0.61 \pm 0.20$,
and for the Verheijen (1997) sample, $\xi=0.58 \pm 0.09$. Such lower
values of $\xi$ are found when a rotation curve rises slowly.

Indeed, the conclusion that the stellar disk cannot be scaled to
explain the inner rise of the observed rotation curve is reached in
many of the dark matter studies of late-type dwarf galaxies, e.g.
DDO~154 (Carignan \& Freeman 1988, Carignan \& Beaulieu 1989), NGC~300
(Puche \etal\ 1990), NGC~55 (Puche \etal\ 1991), NGC~3109 (Jobin \&
Carignan 1990), NGC~5585 (C\^ot\'e \etal\ 1991), DDO~168 (Broeils
1992a), IC~2574 (Martimbeau \etal), NGC~2915 (Meurer \etal\ 1996),
five dwarfs in the Sculptor and Centaurus A groups (C\^ot\'e
\etal\ 1997; see also C\^ot\'e 1995), NGC~5204 (Sicotte \& Carignan
1997), NGC~2976 (Simon \etal\ 2003), and NGC~6822 (Weldrake
\etal\ 2003). Many of these studies have led to the notion that dwarf
galaxies are dominated by dark matter at all radii.

There are, however, also studies that provide a different picture.
Some early mass models for late-dwarf galaxies showed that the stellar
disk could be scaled to explain all of the inner rise of the rotation
curves, suggesting that the dark matter properties of these
galaxies were similar to those of spiral galaxies (Carignan 1985;
Carignan, \etal\ 1988). Similar results were found for the four
late-type dwarf galaxies in the Virgo clusters studied by Skillman
\etal\ (1987), DDO~170 (Lake \etal\ 1990), DDO~105 (Broeils 1992a),
NGC~1560 (Broeils 1992b), most of the late-type dwarf galaxies
presented in van Zee \etal\ (1997), the Large Magellanic Cloud (Kim
\etal\ 1998; Bekki \& Stanimirovic 2009), and most of the galaxies
presented in S99 and Swaters \etal\ (2003).

The spread in results seen in earlier studies raises the question
whether the spread reflects an intrinsic spread in the rotation curve
shapes of dwarf galaxies, which then results in a range of different
mass models, or whether this observed spread is due to other factors
that affect the derived rotation curves.

There are well-known systematic effects that can affect the shape of
the derived rotation curve and can make it appear shallower.  For
example: (i) beam smearing can wash out steep gradients and make the
inner rotation curve appear more shallow (e.g., Begeman 1987;
Paper~III); (ii) rotation curves derived from high-inclination
galaxies are susceptible to underestimated velocities in the inner
parts (e.g., Sancisi \& Allen 1979; Swaters \etal\ 2003); (iii) some
methods of deriving the velocity fields, (e.g., adopting the
intensity-weighted mean velocity of a profile) can lead to an
underestimate of the derived rotation curve (e.g., Bosma 1978;
Paper~I; de Blok \etal\ 2008). In addition, in some cases derived
rotation curves may have been affected by noncircular motions (e.g.,
Rhee \etal\ 2004; Spekkens \& Sellwood 2007; Valenzuela \etal\ 2007).

Given that beam smearing has been ignored in many of the previous
studies, that some of the galaxies in those studies have high
inclinations, and that the majority of those studies used
intensity-weighted mean velocity fields, it is likely that the inner
slopes of the rotation curves in those previous studies have been
underestimated. Because such an underestimate of the inner slope
limits how much the contribution of the stellar disk can be scaled in
a maximum disk fit, an underestimate of the inner slope directly
results in an underestimate of $\xi$.

It is beyond the scope of this paper to review each rotation curve in
the literature in detail and assess whether systematic effects may
have played a role. However, it is useful to consider as an example
the well-studied prototype dwarf galaxy DDO~154. From \HI\ synthesis
observations of this galaxy a slowly rising rotation curve was
derived, and mass modeling indicated that DDO~154 was almost entirely
dominated by dark matter (Carignan \& Freeman 1988; Carignan \&
Beaulieu 1989). Based on the published mass model, we estimate that
$\xi=0.52$, much lower than the values found for our sample. Those
early observations had a resolution of $45''$, and the derivation of
the rotation curve was certainly affected by beam smearing, especially
given the galaxy's inclination of approximately $66^\circ$ (de Blok
\etal\ 2008).

Recently, de Blok \etal\ (2008) presented higher resolution \HI\ data
for this galaxy. Because of the $12''$ resolution, and the fact that
the velocity field was derived using skewed Gaussians, which can
account for asymmetries often observed in line profiles, the derived
rotation curve is less prone to the systematic effects discussed
here. The rotation curve derived by de Blok \etal\ (2008) rises
somewhat more steeply than the one derived earlier by Carignan \&
Beaulieu (1989), and de Blok \etal\ (2008) find $\xi=0.62$.  However,
the position-velocity diagram shown in Figure~81 of de Blok
\etal\ (2008) suggests that the rotation curve in the inner $2'$ could
be even steeper. Indeed, a new study of the kinematics of DDO~154 in
the central regions based on long-slit spectroscopy and integral-field
spectroscopy (Swaters \etal, in preparation) shows a rotation curve
that rises somewhat more steeply than the one of de Blok
\etal\ (2008), resulting in a value of $\xi=0.80$.

A key feature of the sample presented in this paper is that the
systematic effects mentioned above were avoided as much as
possible. The sample galaxies have inclinations between $39^\circ$ and
$80^\circ$ and galaxies with large asymmetries were excluded. The
rotation curves were derived with an interactive technique using
modeling of the observed data cube to correct for the effects of beam
smearing (Paper~III). 

\subsection{Mass to light ratios of the stellar disk}
\label{secmls}

For the maximum disk models, the derived mass to light ratio
$\mlstarR$ is tightly coupled to surface brightness, with lower
surface brightness galaxies having higher $\mlstarR$. This is expected
in order to explain the high mass density needed for the maximum disk
despite the LSB nature of the stellar disks.  The values for
$\mlstarR$ required in the maximum disk models, reach as high as
$\mlstarR=15$, well outside the range of what population synthesis
models predict (e.g., Bell \etal\ 2003; Zibetti \etal\ 2009). The
precise value of $\mlstarR$ as derived from such population synthesis
may be uncertain by a factor of 2 to 4 (Portinari \etal\ 2004), and
possibly even higher (Bershady \etal\ 2010). However, even considering
these uncertainties, values as high as $\mlstarR=15$ appear unlikely,
as they would require a strong bias in the initial mass function
towards low-mass stars, or a large fraction of the stellar mass being
locked up in stellar remnants.

The presence of a nonstellar component of mass with a distribution
similar to that of the stars, e.g., in the form of cold molecular gas
as suggested by Pfenniger \& Combes (1994) and Revaz \etal\ (2009),
could provide an explanation.

Alternatively, the stellar disks may be somewhat less than maximal.
Independent measurements of \mlstar\ from stellar velocity dispersions
in spiral galaxies suggest that stellar disks have a submaximum
contribution to the rotation curve, with a peak contribution of the
stellar disk to the rotation curve of around 65\% (Bottema 1993;
Westfall \etal, in preparation). At this level, the stellar disk still
determines the inner shape of the rotation curve, i.e., one would
still expect to find $\xi\sim 1$. If the stellar disks would also
contribute around 65\% in late-type dwarf galaxies, this would reduce
$\mlstarR$ by almost a factor of 2.5. This would bring the mean and
maximum $\mlstarR$ of the galaxies in our sample to 3 and 6,
respectively, still above what population synthesis models predict.

If $\mlstarR$ is assumed to be near unity, as predicted by stellar
population synthesis models (see Section~\ref{subrotcurfits}), then
well-fitting mass models are found as well. In this case, all galaxies
except UGC~7577 are dominated by dark matter. Moreover, the relative
contribution of dark matter increases towards lower surface
brightness, and the contribution of the disk decreases
correspondingly. Thus, for models with $\mlstar\sim 1$, we reach the
same conclusions as are usually found for dwarf and LSB galaxies
(e.g., Carignan \& Beaulieu 1989; Broeils 1992a; de Blok \& McGaugh
1997).

There is, however, a critical difference. In many of those previous
studies, as described above, low values for $\xi$ were derived. For
our sample, however, we find a value for $\xi$ that is near unity.
This not only means that the stellar disk can be scaled to explain
most of the inner parts of the observed rotation curve. If
$\mlstarR\sim 1$, this also means that the rotation curve of the
stellar disk alone must have the same shape as that of the dark matter
that dominates the potential. Otherwise, $\xi$ could not be near
unity.

In spite of the uncertainties in $\mlstarR$, one result from
this analysis is firm: the fact that the contribution of the stellar
disk can be scaled up to explain the inner parts of the rotation curve
indicates that the shapes of the rotation curve of the stellar disk
and of the observed rotation curve are similar. There appears to be a
strong coupling between the distribution of the stars and that of the
total mass (see also Sancisi 2004; Paper~III).

\section{Conclusions}
\label{theconclusions}

We have presented mass models for a sample of 18 late-type dwarf
galaxies. For this sample, the systematic effects due to beam
smearing, intensity-weighted-mean velocity fields, high inclination,
and noncircular motions were avoided as much as possible.  We have
analyzed the rotation curves of the galaxies in our sample by
fitting mass models consisting of a stellar disk, a gaseous disk and
the contribution of an isothermal halo, and we have reached the
following conclusions:

\noindent 
(1) Dwarf galaxies are not necessarily dominated by dark matter within
their optical radii.  The rotation curve of the stellar disk can be
scaled up to explain most of the inner parts of the observed rotation
curves.  The maximum disk mass models are similar to those for spiral
galaxies, with similar fraction of dark matter within a radius of four
disk scale lengths. However, the required stellar mass-to-light ratios
are high, up to about 15 in the $R$-band.

\noindent
(2) For the maximum disk models, the contribution of the stellar disk
to the rotation curve is about 90\% at two disk scale lengths,
independent of surface brightness or luminosity. The maximum-disk
\mlstar\ is correlated with surface brightness.

\noindent
(3) Well fitting mass models can be obtained by assuming that the
contribution of the stellar disk to the rotation is maximal, but also
by assuming it is negligible. This demonstrates
that mass models for late-type dwarf galaxies suffer from the same
indeterminacies as those for spiral galaxies.

\noindent
(4) If the stellar mass-to-light ratios are set to the values
predicted by stellar population synthesis modeling, the relative
contribution of the stellar disk decreases with surface brightness,
while the relative fraction of dark matter increases.

\noindent
(5) Whatever the contribution of the stellar disk to the rotation
curve, the similarity in shapes between the rotation curve expected
for the stellar disk and the observed rotation curve implies that in
the inner parts of late-type dwarf galaxies the total mass density
distribution is closely coupled to the luminous mass density
distribution.

\clearpage

\clearpage

\clearpage

\clearpage

\clearpage

\end{document}